\documentclass[twocolumn]{aastex63}
\usepackage{xcolor}

\usepackage{color, colortbl}
\definecolor{Gray}{gray}{0.75}
\usepackage{graphicx}
\usepackage{amssymb}
\usepackage{amsmath}
\usepackage{relsize}
\usepackage{url}
\usepackage{color, colortbl}
\usepackage{enumerate}  
\usepackage{appendix}

\begin{document}

\title{Origin and extent of the opacity challenge for atmospheric retrievals of WASP-39~b}

\author[0000-0002-8052-3893]{Prajwal Niraula}
\affiliation{Department of Earth, Atmospheric and Planetary Sciences, MIT, 77 Massachusetts Avenue, Cambridge, MA 02139, USA}

\author[0000-0003-2415-2191]{Julien de Wit}
\affiliation{Department of Earth, Atmospheric and Planetary Sciences, MIT, 77 Massachusetts Avenue, Cambridge, MA 02139, USA}

\author[0000-0003-4763-2841]{Iouli E. Gordon} 
\affil{Harvard-Smithsonian Center for Astrophysics, Atomic and Molecular Physics Division, Cambridge, MA, USA}

\author[0000-0002-7691-6926]{Robert J. Hargreaves} 
\affil{Harvard-Smithsonian Center for Astrophysics, Atomic and Molecular Physics Division, Cambridge, MA, USA}

\author[0000-0002-7853-6871]{Clara Sousa-Silva}
\affil{Physics Program, Bard College, 30 Campus Road, Annandale-On-Hudson, New York 12504, USA}
\affil{Centro de Astrofísica da Universidade do Porto,  Portugal}

\begin{abstract}
As the James Webb Space Telescope (JWST) came online last summer, we entered a new era of astronomy. This new era is supported by data products of unprecedented information content that require novel reduction and analysis techniques. Recently, Niraula et al. 2022 (N22) highlighted the need for upgraded opacity models to prevent facing a model-driven accuracy wall when interpreting exoplanet transmission spectra. Here, we follow the same approach as N22 to explore the sensitivity of inferences on the atmospheric properties of WASP-39~b to the opacity models used. We find that the retrieval of the main atmospheric properties from this first JWST exoplanet spectrum is mostly unaffected by the current state of the community’s opacity models. Abundances of strong opacity sources like water and carbon dioxide are reliably constrained within $\sim$0.30~dex, beyond the 0.50~dex accuracy wall reported in N22. Assuming the completeness and accuracy of line lists, N22's accuracy wall is primarily driven by model uncertainties on broadening coefficients and far-wing behaviors, which we find to have marginal consequences for interpreting the transmission spectra of large, hot, high-metallicity atmospheres such as WASP-39~b, in opposition to emission spectra and climate modeling which depend on deeper parts of a planetary atmosphere. The origin of the opacity challenge in the retrieval of metal-rich hot Jupiters via transmission spectroscopy will thus mostly stem from the incompleteness and inaccuracy of line lists.
\end{abstract}


\section{Introduction} 
\label{sec:intro}

WASP-39~b is a canonical puffy Saturn-sized planet \citep{Faedi2011} whose atmospheric characterization prior to JWST had already shown prominent atmospheric features. \citet{wakeford2018} observed strong water absorption features with Hubble observations using the Wide Field Camera with hints of carbon dioxide when combined with the Spitzer observations. The low-significance detection of CO$_2$ could not fully distinguish between sub or super-solar metallicity, remaining a point of contention in the subsequent studies as a range of C/O ratios were obtained \citep{wakeford2018, tsiaras2018, fisher2018, pinhas2019, welbanks2019, min2020}. Fortunately, the recent high-precision JWST observation provided definitive evidence for a strong CO$_2$ absorption feature in the transmission spectrum of WASP-39~b \citep[][henceforth \texttt{JTECERST23}]{JWST_ERS_WASP39}. 

The current interpretation of WASP-39~b's spectrum has been performed via a direct comparison of different forward models to the data. These comparisons suggest varying C/O ratios anywhere from 0.23 to 0.7 (see \texttt{JTECERST23}). \citet{niraula2022_Nature} [henceforth \texttt{N22}] recently showed that the sensitivity of atmospheric retrieval outputs combined with the current limitations of opacity models will result in a 0.5 to 1 dex accuracy wall in our interpretation of the new generation of exoplanetary spectra. The main contributors to this opacity-driven accuracy wall appear to be the treatment of broadening and far-wings, if one assumes that the line lists are complete and accurate. As the precision of this first JWST exoplanet transmission spectrum already appears to contain enough information to warrant concerns regarding the aforementioned opacity-driven accuracy wall, we chose to explore the impact of the state of current opacity models on its interpretation with a particular focus on the interaction between CO$_2$ and H$_2$.

\begin{figure}[th!]
\centering
\includegraphics[trim={0.2cm 2.0cm 0.2cm 1.0cm},clip, angle=0, width=0.45\textwidth ]{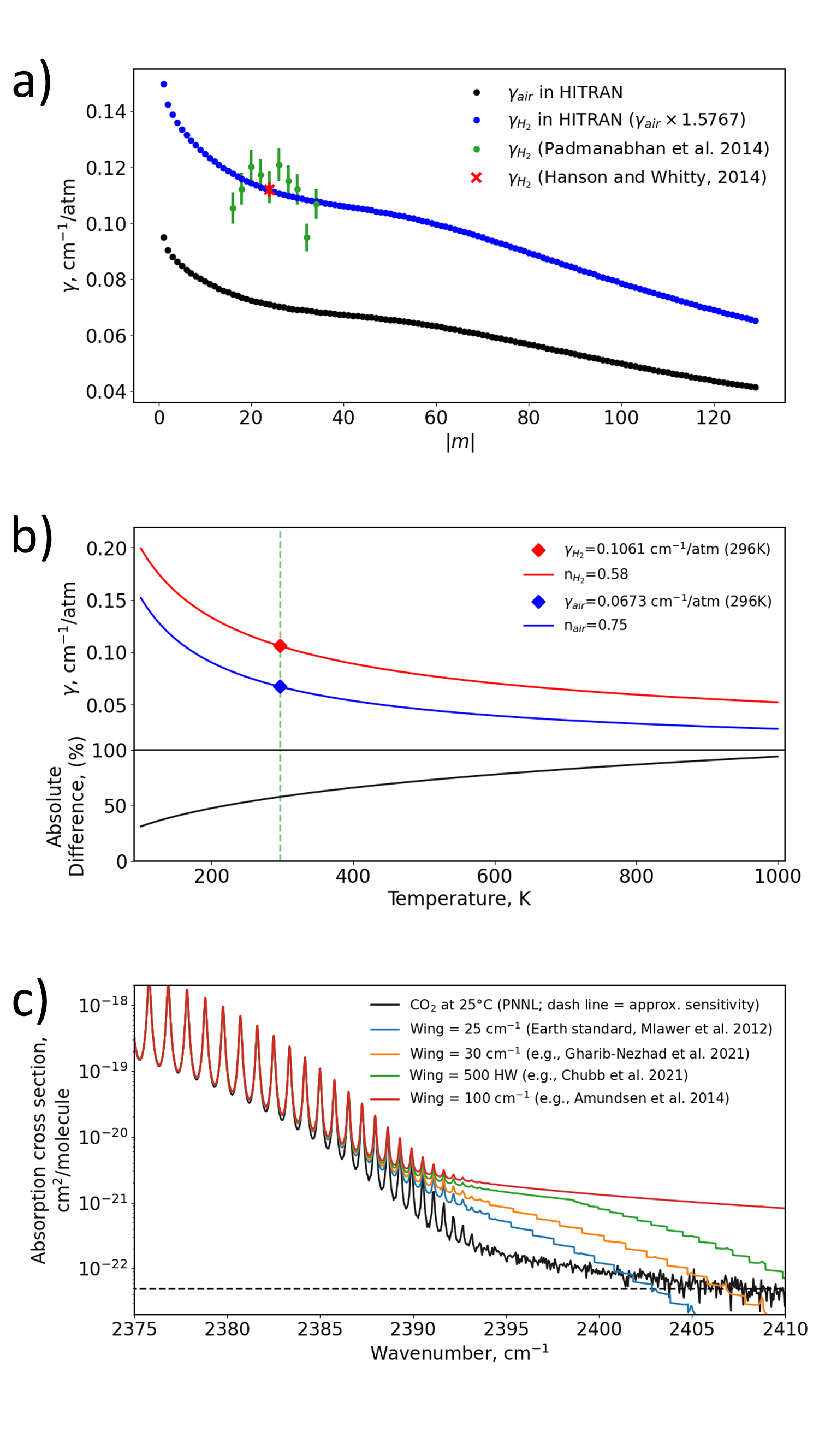}
\figcaption{An overview of the limitations of spectral line parameter for CO$_{2}$. a) The sparse data for the CO$_{2}$-H$_{2}$ system. Here, experimentally measured half-widths \citep{padmanabhan2014, hanson2014} do not provide sufficient coverage for populating the HITRAN database \citep{hitran2020}. The CO$_{2}$-air broadening parameters in HITRAN have therefore been scaled to the measurement of \citet{hanson2014} to retain rotational structure \citep{tan2022}, even though rotational dependence is not expected to be exactly the same. b) The effect of temperature dependence. A demonstration of the combined effect of using the CO$_{2}$-H$_{2}$ and CO$_{2}$-air parameters (i.e., $\gamma$ and $n$)  for a typical $P(40)$ transition in HITRAN. In this example, the relative difference increases at higher temperatures. c) Limitations of the Voigt profile regarding the far-wing behavior of CO$_{2}$ lines. Experimental absorption of the 4.3~$\mu$m at 25$^{\circ}$C in 1~atm N$_{2}$ \citep[PNNL,][]{PNNL2004} compared to modeled spectra that use HITRAN2020 parameters with different line wing cut-offs that are typically used for Earth \citep{mlawer2012} and exoplanetary atmospheres \citep{amundsen2014, chubb2021, GharibNezhad2021}. The approximate experimental sensitivity is shown as the dashed line.   \label{fig:BroadTempWing}}
\end{figure}

The CO$_2$-H$_2$ system will be central to the atmospheric exploration of worlds accessible with JWST. The vast majority of them will showcase the CO$_2$ 4.3-$\mu$m absorption feature broadened in large part by H$_2$, the dominant atmospheric constituent of puffy atmospheres. Yet, the knowledge available for the CO$_2$-H$_2$ system is very limited (Table 7 of \citet{tan2022}). Only extremely sparse measurements exist \citep{padmanabhan2014, hanson2014}, which were used to scale a fairly complete dataset of air-broadened half-widths of CO$_2$ lines \citep{hashemi2020} in HITRAN \citep{hitran2020}--see Figure\,\ref{fig:BroadTempWing}. The aforementioned limitations are even more pronounced for the far-wing effect, which is one of the least studied and understood areas of molecular spectroscopy, despite its important role in radiative transfer \citep{hartmann2018}. As an example, the far wings of H$_{\rm 2}$O, which are major contributors to the water continuum \citep{mlawer2012}, are parametrized differently than those of CO$_2$ lines that are assumed to be sub-Lorentzian \citep{cousin1985}. For different molecules and their possible collisional broadeners, the difference in far-wing behavior can be drastic (e.g., the CH$_4$-H$_2$ system in \citet{hartmann2002}). For terrestrial atmospheres, a Voigt line wing cutoff of 25 cm$^{-1}$ is typically used \citep{mlawer2012}. For the aforementioned water vapor and carbon dioxide empirical continuum models are being implemented to account for far-wing as well as other effects resulting in continuum-like absorption \citep{mlawer2012}. The common practice for exoplanet atmospheres is that the same Voigt profile is used but with higher cutoff values, such as 30 cm$^{-1}$ \citep{GharibNezhad2021},  500 half-widths \citep{chubb2021}, and 100 cm$^{-1}$ \citep{amundsen2014}. These are very different approaches that do have an impact where the densities of ambient gases (perturbers) are sufficiently high. 

In this study, we investigate the extent to which the interpretation of the JWST/NIRSpec/G395H transmission spectrum of WASP-39\,b is sensitive to limitations stemming from current opacity models. This study will contextualize the findings of N22 and further inform us of the possible urgency and relevance of efforts aiming at improving such models. To this end, we follow the same framework as N22 and perform a perturbation/sensitivity analysis of the inferences gained about WASP-39~b's atmospheres via cross-retrieval using different opacity models. In Section \ref{sec:data}, we briefly introduce the data used. In Section \ref{sec:analysis}, we present the framework, the different cross-sections supporting our sensitivity analysis as well as our atmospheric models. We introduce the results of our analysis in Section \ref{sec:results} and discuss them while considering large implications for JWST's optimal use in Section \ref{sec:implications}.

\begin{figure*}[t!]
\centering
\includegraphics[trim={0cm 0cm 0cm 0cm},clip, angle=0, width=0.95\textwidth ]{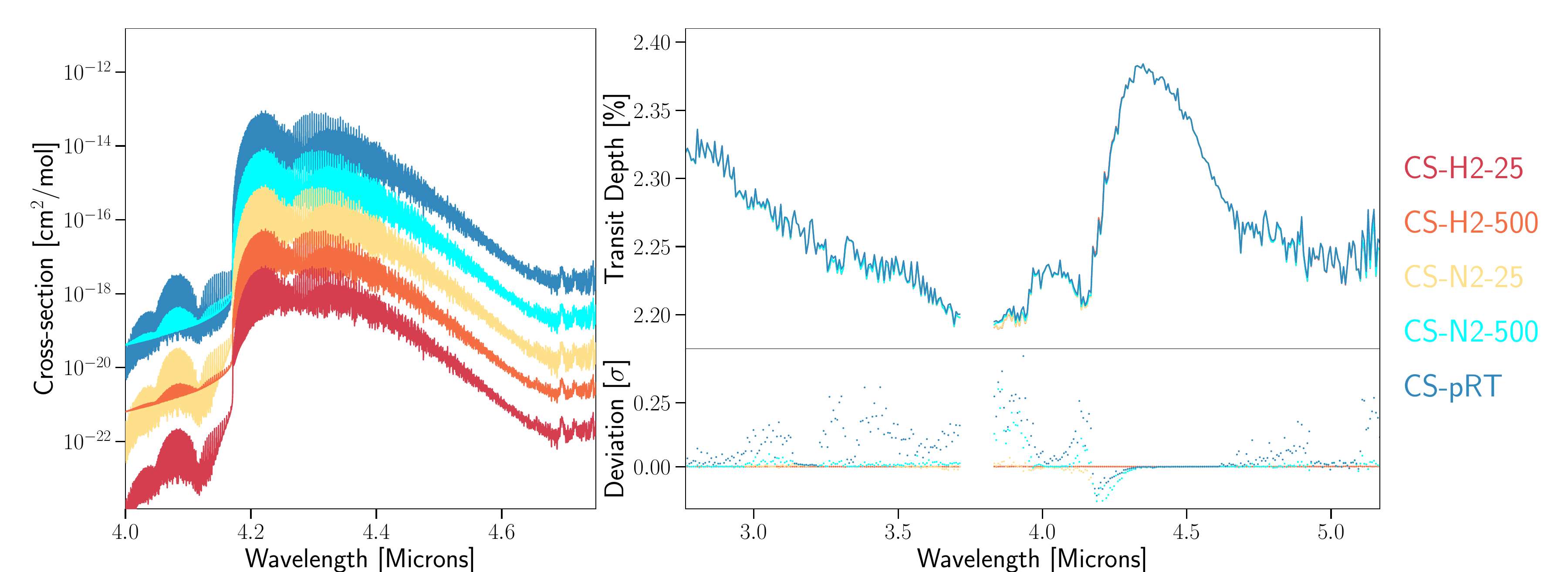}
\figcaption{\textbf{Left}: Comparison of carbon dioxide 4.3-micron feature closest to the temperature of 900 K and 1 atm pressure for the five opacity models used in our study. The different opacity models are offset for the purpose of clarity. \textbf{Top Right:} Comparison of models at the observational resolution of NIRSpec/G395H generated for WASP-39b using five different cross-sections using mass of 0.28 M$_{\rm Jup}$, P$_{\rm 0}$ 1.0 atm, T$_{\rm 0}$ 900.0 K, Log MR$_{\rm CH_4}$ -5.0, Log MR$_{\rm CO}$ -10.0, LogMR$_{\rm CO_2}$ -2.0, Log MR$_{\rm H_2O}$ -1.0, Log MR$_{\rm H_2S}$ -4.0,
LogMR$_{\rm HCN}$ -4.0, and Log MR$_{\rm SO_2}$ -4.0. \textbf{Bottom Right:} The observed deviations obtained from the inter-model comparisons. The transmission model using CS-H2-25 was used as the base model for comparison. \label{fig:combined_figure1}}
\end{figure*}

\section{Data}
\label{sec:data}
WASP-39b was selected as part of the Director’s Discretionary-Early Release Science (ERS-1366) to demonstrate the capabilities of the various spectroscopic JWST instruments and to understand their systematics. Among these instruments, NIRSpec/G395H has one of the highest resolutions and wide spectral coverage.  For the observation, the 2.8 hours transit of WASP-39~b was recorded with two detectors: NRS1 covering 2.725 to 3.716 microns and NRS2 detector covering 3.829 to 5.172 microns at a resolution of $\sim$600. The observations were taken on 30-31 July 2022 between 21:45 -06:21 UTC and suffered from a tilt event. This could be corrected, and the data were within 2$\times$ of the photon noise limit. This paper uses the publicly available data reduced light curve made available through 10.5281/zenodo.7185300 \citep{alderson2023}. The data were reduced using 11 individual pipelines and combined to create a weighted average model. For more details about the data reduction and other systematics, please refer to \citet{alderson2023}.

\section{Analysis Framework}
\label{sec:analysis}

We use the framework introduced in \texttt{N22}\footnote{Framework publicly available at \\ \href{https://github.com/disruptiveplanets/tierra}{https://github.com/disruptiveplanets/tierra}.} to analyze the JWST/NIRSpec/G395H
transmission spectrum of WASP-39\,b. The essence of the framework is to perform retrievals with a selection of cross-sections that have been chosen to represent limitations (incl., uncertainties, and incompleteness) of the underlying line lists or calculations. Differences in the retrieved atmospheric parameters will therefore highlight uncertainties introduced by such limitations. The selected cross-sections include perturbations to line positions, line intensities, or line broadening based on their uncertainties, or incorporate different treatments to the cross-section calculations present in the literature (e.g., line wing limit). Among the major differences from \texttt{N22}, we focus on hydrogen/helium-dominated atmospheres and follow non-isothermal temperature profiles parametrization of \citet{madhusudhan2009}.

\subsection{Opacity Models}
\label{sec:cross_sections}

We generate representative cross-sections for air-broadened and hydrogen-broadened cross-sections for seven different molecules particularly relevant for the retrieval of  WASP-39~b (Figure\,\ref{fig:combined_figure1}). H$_{\rm 2}$O, CO$_{\rm 2}$ and SO$_{\rm 2}$ have been detected with significance in \texttt{JC},  at the 21.5-, 28.5-, and 4.8-$\sigma$ levels, respectively. We generate opacity cross-sections for the main isotopologues of these three molecules as well as CO, H$_{\rm 2}$S, HCN, CH$_{\rm 4}$, and SO$_{\rm 2}$. For H$_{\rm 2}$O, CO$_{\rm 2}$, CO and CH$_{\rm 4}$, HITEMP line lists are used \citep{HITEMP2010, Li2015, hargreaves2020} because of the high temperature of the atmosphere. The HITRAN line lists \citep{hitran2020} are used for H$_{2}$S, HCN, and SO$_{\rm 2}$, because they are not available in HITEMP. In general, the use of HITRAN is not recommended at the temperature relevant to the atmosphere of WASP-39~b, as for most of the molecules, it is complete only at temperatures relevant to the terrestrial atmosphere. Therefore, it may not be able to accurately reproduce high-temperature spectra even at lower resolution. With that being said, validations of the SO$_2$ line list in HITRAN2020 against high-resolution laboratory spectra at temperatures up to 800K, proved that it was sufficiently accurate for the band centers. Nevertheless, the temperature limitation of the HITRAN lists should be noted. Moreover, in the case of SO$_2$ and H$_2$S, the combined contribution of absorption from the isotopologues to the total band intensity should be about 6\%. For this study, the following five opacity models were generated with the aim of performing a sensitivity analysis of WASP-39\,b:

\begin{enumerate}[(i)]
     \item \textbf{CS-H2-25}: This cross-section represents a typical hydrogen/helium molar ratio of 0.15 \citep{conrath1987}. We use the Voigt model for the line profile, and truncate the lines at 25 cm$^{-1}$ as prescribed for \texttt{HITRAN} \citep{hitran2020}. We use the same pressure-temperature (p-T) grid as in \texttt{N22} designed specifically to decrease the interpolation error.  The hydrogen and helium broadening parameters are taken from the HITRAN database, whenever available. 
    
    \item \textbf{CS-H2-500}: This cross-section aims to understand the impact of the line wings in transmission spectroscopy. We extend the line cut-off value to 500 cm$^{-1}$ otherwise is equivalent to CS-H2-25. The line wings, for the strong lines often determine the opacity continuum, particularly at higher pressures.
    
    \item \textbf{CS-N2-25}: This cross-section aims to look at the impact of the choice of the broadening parameter i.e. using air broadening values instead of hydrogen. This cross-section is produced with typical parametric choices used in \texttt{hapi}\footnote{\href{https://github.com/hitranonline/hapi}{https://github.com/hitranonline/hapi}}, a python code to calculate cross-sections from HITRAN/HITEMP data. Typically, air-broadening parameters are larger than hydrogen/helium-broadening parameters, with some exceptions, including CO$_2$ lines as shown in Figure \ref{fig:BroadTempWing}(a). 

    \item \textbf{CS-N2-500}: This cross-section aims to study the combined effect of broadening and line wings on transmission spectroscopy. For this set of cross-sections, we generate the cross-section up to 500 cm$^{-1}$ with air-broadening parameters. Otherwise, it is equivalent to CS-N2-25.  
    
    \item \textbf{CS-pRT}: We use the cross-section provided for \texttt{petit-RADTRANS} \citep{petitRadtrans2019} and adapt it to work with \texttt{tierra}\footnote{\href{https://github.com/disruptiveplanets/tierra}{https://github.com/disruptiveplanets/tierra}}. These cross-sections use air broadening parameters, and the line wings are treated with the formulation from \citet{hartmann2002}.  The temperature-pressure grid is 3-5$\times$ coarser than the one used in \texttt{N22} and extends up to $\sim$3000 K. When the cross-sections were missing, such as in the case of sulfur dioxide, we fall back to the CS-H2-25.  We reduced the cross-section to a resolution of 100,000 from 1 million using linear interpolation.
\end{enumerate}

There is no consensus for the p-T grid to be used in exoplanet retrievals and we have chosen \texttt{petit-RADTRANS}  for this study due to its public availability and high resolution. The grid introduced in \texttt{N22} led to an error around ~3\% with the bilinear interpolation scheme, which is marginal for JWST applications. A sparser grid is thus expected to increase the contributions of model-driven uncertainties to the overall uncertainty budget. 

\subsection{Atmospheric Model}
\label{sec:model}
Our retrieval code \texttt{tierra} assumes volume iso-mixture profile throughout the height of the atmosphere.  For all of our cross-sections, a native resolution of 100,000 was chosen to ensure a negligible effect on the retrieval performance considering NIRSpec and the precision of the data  (see \texttt{N22}). Our retrievals constrain the abundances of seven different molecules using \texttt{tierra} with the nominal and perturbed cross-sections. In order to explore the solution space of the parameters, we run MCMC using \texttt{emcee} \citep{emcee} for 50,000 steps with four times the number of walkers as the number of parameters. Once the fits have converged, we remove the burn-in part of the chains to build the posterior distribution of the parameters. Altogether we have 11 parameters for isothermal models, and 13 parameters for parametric p-T models. For our retrievals models for WASP-39 b, we consider the aforementioned seven molecules with volume mixing ratio priors between 0.5 to -25.

\begin{figure*}[ht!]
\centering
\includegraphics[trim={0cm 2cm 0cm 0cm},clip, angle=0, width=0.95\textwidth ]{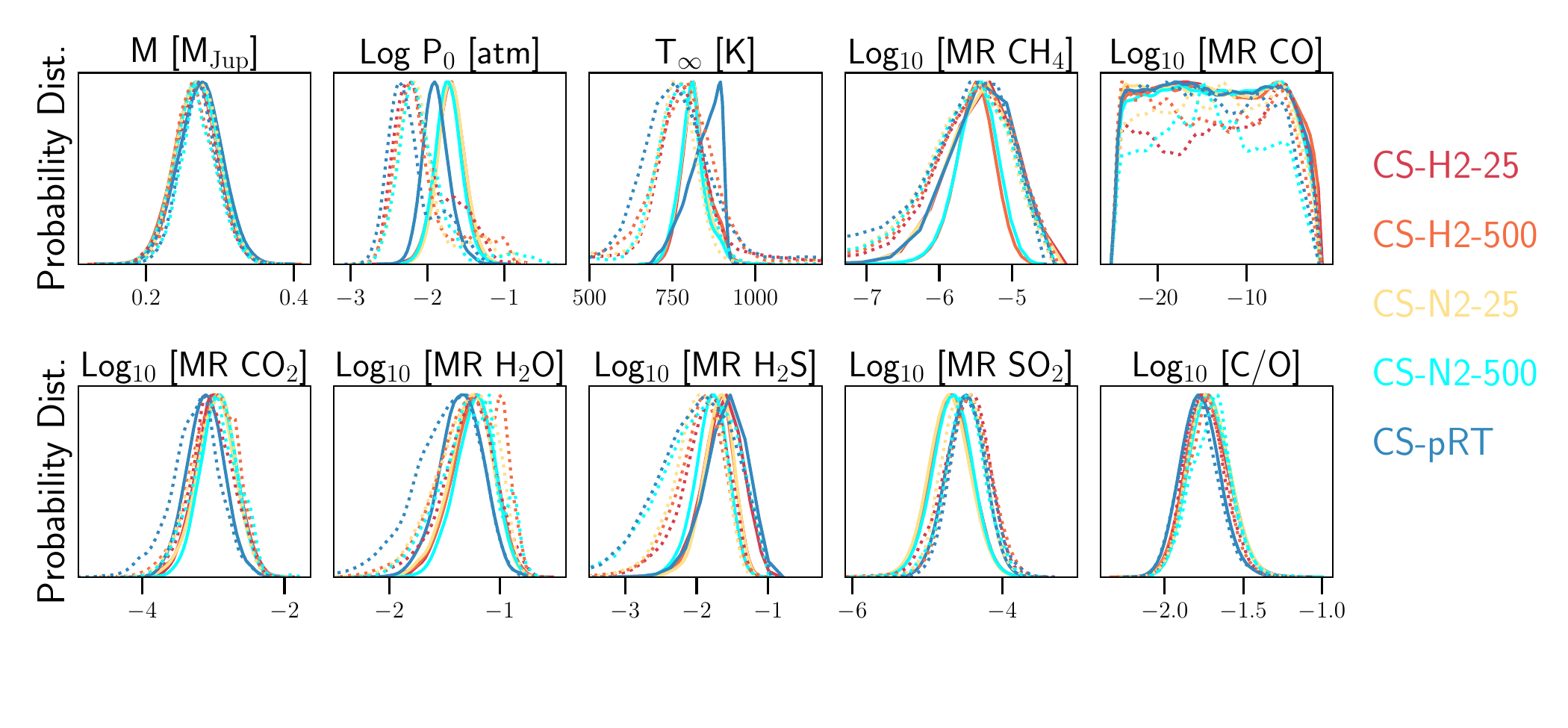}
\figcaption{ Posterior distribution of the retrieved parameters retained for WASP-39 b for NIRSpec G395H for the isothermal (solid) and parametric (dotted) p-T atmospheric model. The posteriors of the parameters are tabulated in \autoref{tab:retrieved_abundances_wasp39}. The scatter among the different retrievals when using different cross-sections is inherently larger than the observed precision for different molecules. The right most panel of the bottom row shows carbon to oxygen ratio calculated from the retrieved carbon and oxygen bearing molecular species.} \label{fig:ppd_combined}
\end{figure*}

\section{Results}
\label{sec:results}

\autoref{fig:ppd_combined} shows the posterior distribution of the retrieved atmospheric parameters with the corresponding values tabulated in \autoref{tab:retrieved_abundances_wasp39}. Comparable to \texttt{JTECERST23}, we find a high abundance of H$_{\rm 2}$O, CO$_{\rm 2}$, and H$_{\rm 2}$S. We note that our retrievals yield a C/O ratio tenfold lower (log$_{\rm 10}$ [C/O] $\sim$ -1.75, 30$\times$ solar metallicity based on the atmospheric composition) compared to reported values in \texttt{JTECERST23}, likely due to the fact that we do not introduce priors based on the chemical models (\texttt{JTECERST23} assumes 1D radiative-convective thermochemical equilibrium). Most importantly, we find that all tested opacity models yield retrieved atmospheric parameters within the error bars. This implies that the limitations due to broadening considerations have marginal effects on the analysis and interpretation of one NIRSpec/G395H transmission spectrum of WASP-39~b. It also implies that main atmospheric properties (i.e., associated with dominant absorbers and p-T profile) can reliably be constrained beyond the 0.5~dex opacity-driven accuracy wall reported in N22. For example, the abundance of strong opacity sources such as H$_{\rm 2}$O and CO$_{\rm 2}$ can be reliably constrained in this particular environment within $\sim$0.3 dex.

\vspace{-0cm}\begin{figure}[t!]
\hspace{-1.3cm}
\centering
\includegraphics[trim={0.25cm 0.0cm 0cm 2cm},clip, angle=0, width=0.55\textwidth ]{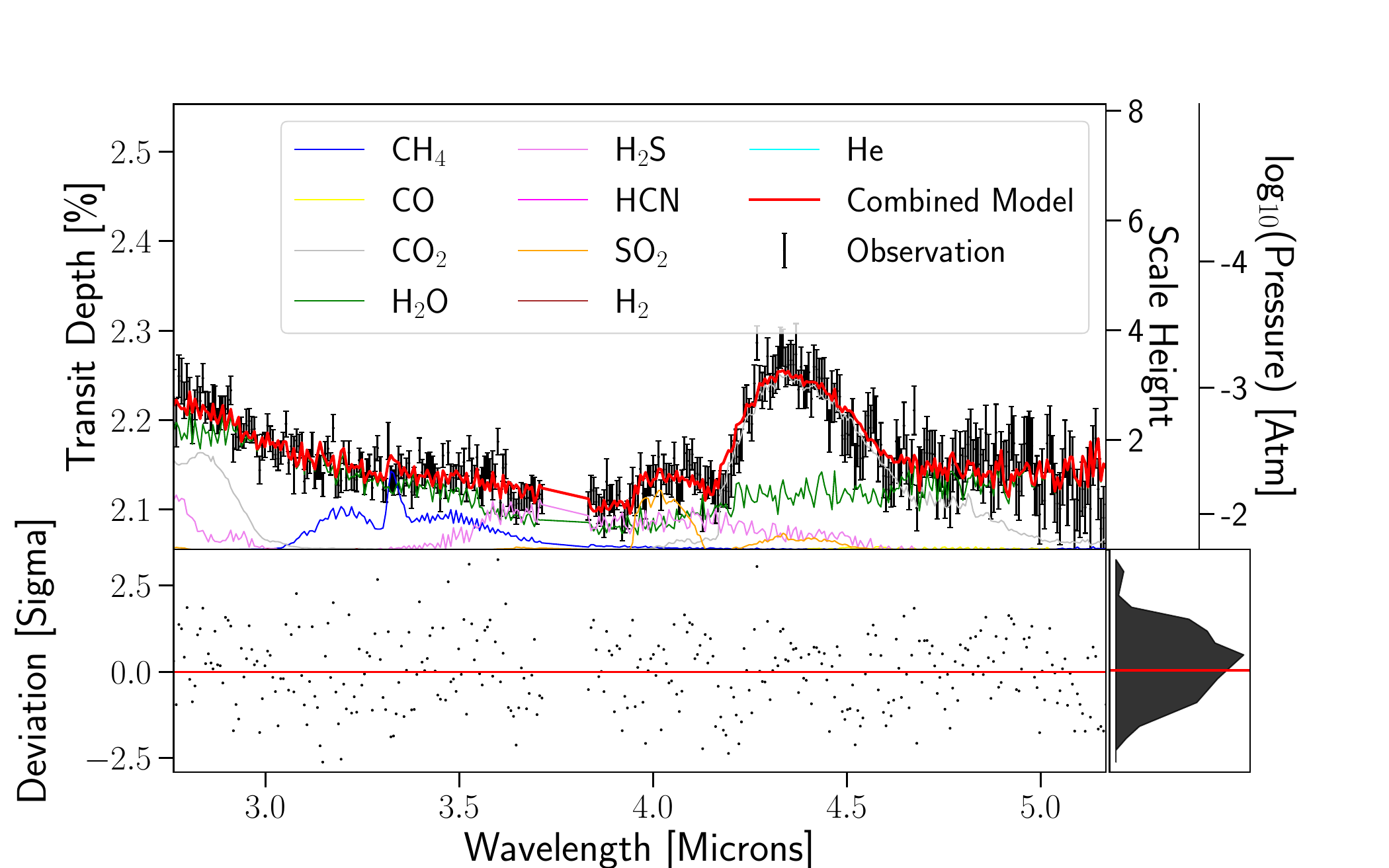}
\figcaption{Best fit showing different molecules using the results from the CS-H2-25 cross-sections. The isothermal model yields  $\chi^2_{\nu}$ of 1.08 showing the adequacy of the current model to explain the trends observed in the data.  \label{fig:BestFitFigure}}
\end{figure}

We also note that the parametric p-T model improves the overall fit and mildly improves the information criteria (AIC \& BIC) compared to the isothermal profile (see \autoref{tab:retrieved_abundances_wasp39}), which in itself provides a good fit (see \autoref{fig:BestFitFigure}). Although the p-T models yield consistent results (mostly within their 1$\sigma$ error bars), the small differences between the reported values highlight the sensitivity of the inferences to the p-T models resulting notably in a final accuracy on the abundance of H$_2$O and CO$_2$ of $\sim$0.3~dex rather than $\sim$0.2~dex.

\section{Discussion}
\label{sec:implications}

Our results may appear in tension with the results reported in N22 as we observe the retrieved parameters are independent of the opacity model choices, and thus no apparent ``opacity-driven accuracy wall'' stemming from pressure broadening effects (notably, broadening coefficients and far-wing parameterization). While the cross-sections considered in this study do not fully explore the issues with completeness e.g. HITRAN vs HITEMP vs ExoMol, the broadening parameters are not perturbed to the same extent, the  precision of the data is 10$\times$ lower, and  the wavelength range spans 5$\times$ smaller, the results are consistent at a level that is better than 0.5 dex, an accuracy threshold identified in the earlier study.  We explore below the possible reasons for this apparent discrepancy. 

\begin{figure}[ht!]
\hspace{-1cm}
\includegraphics[trim={0cm 0.25cm 0cm 1cm},clip, angle=0, width=0.55\textwidth ]{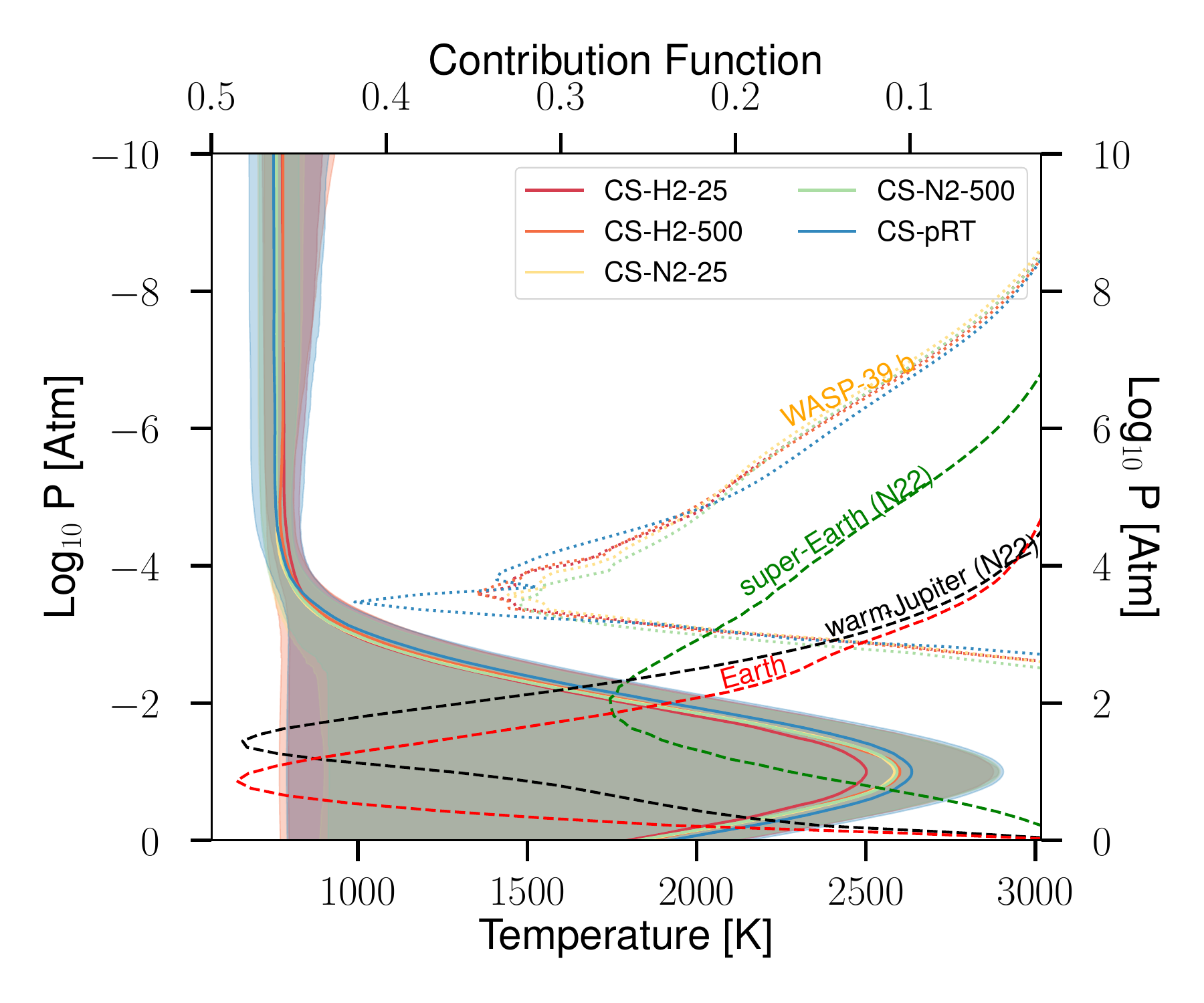}
\figcaption{\label{fig:PT_Profile}p-T profile for five different cross-section retrievals constructed and the corresponding 1$\sigma$ confidence interval. The contribution function as a function of pressure is also shown with axis going from right to left. The dotted lines along the y-axis show the pressure levels most probed for the present observations of WASP-39~b. The two synthetic cases from N22 are shown in dashed lines--a Super-Earth (green), and a warm Jupiter (black), meanwhile, the Earth is shown in dashed red. } 
\end{figure}

First, we explore the SSD (Spectral Statistical Distance) metric used in N22 to quantify the effect of opacity-model perturbations on a planet's transmission spectrum. The SSD for WASP-39~b's NIRSpec/G395H spectrum is much lower than reported for the synthetic cases in N22 ($\lesssim$1 vs $\geq5$), for example, 0.17 for CS-H2-25 compared to CS-H2-500 (see \autoref{tab:retrieved_abundances_wasp39} \&  \autoref{fig:combined_figure1}). To expand on this crucial difference, the impact of  pressure-broadening perturbations is prominent if the radiative properties of the atmospheric layers probed by the transmission spectrum of a planet are affected by pressure. Typically, the pressure regime above the $\sim$10~mbar cannot be modeled via a pure Doppler approximation \citep[][Figure S.2.]{dewit2013}. We compare in  \autoref{fig:PT_Profile} the atmospheric contribution functions for WASP-39~b, the synthetic planetary scenarios used in N22, and Earth. It highlights the pressure level probed in WASP-39~b's atmosphere is close to two orders of magnitude lower than the cases used in N22, specifically $\sim$0.1~mbar, and thus mostly insensitive to pressure-related effects.

\vspace{0cm}

\begin{figure*}[t!]
\centering
\includegraphics[trim={6cm 0.0cm 1cm 0.25cm},clip, angle=0, width=0.98\textwidth ]{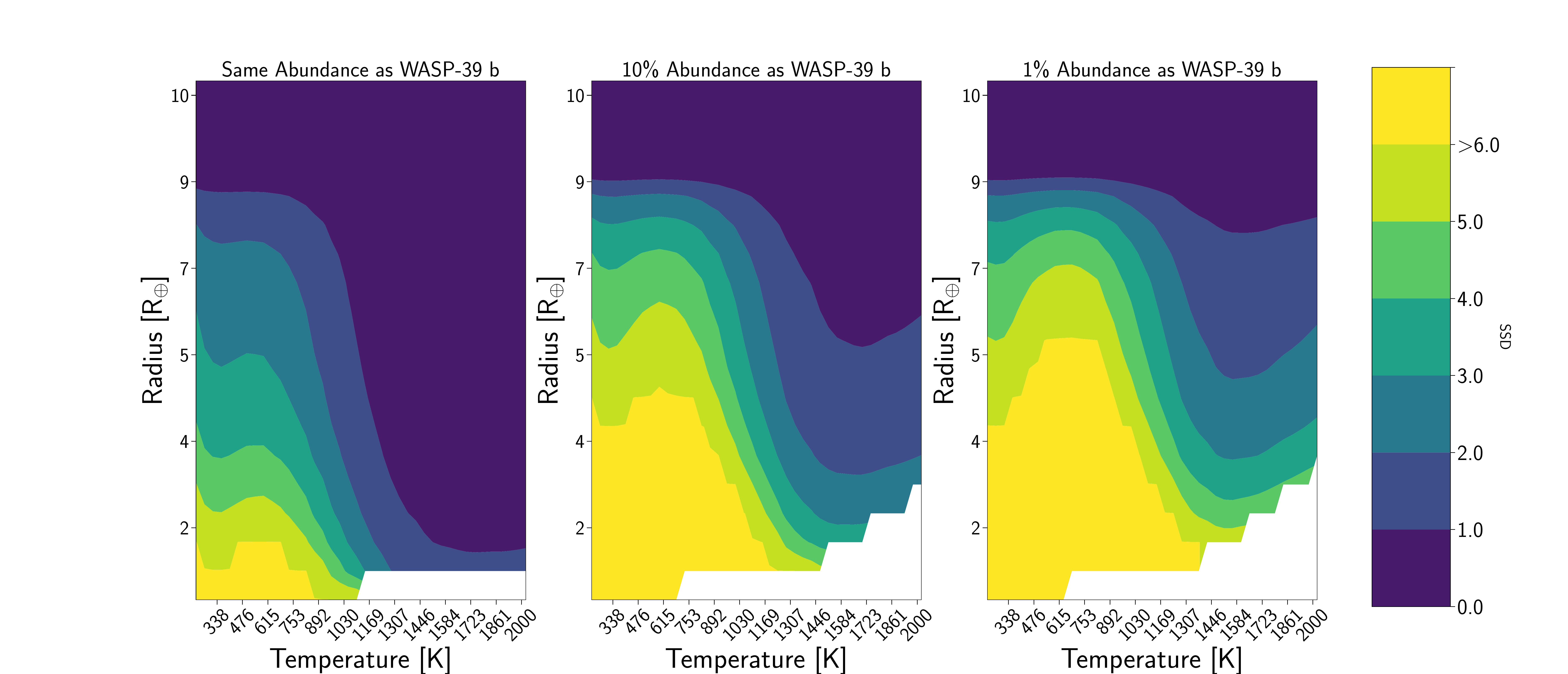} 
\figcaption{\textbf{Left:} The change of SSD with temperature and the size of the planet with 5 JWST transits with the same observational noise and abundance as WASP-39\,b. We assume the planet mass follows the relationship from \citet{chen2017} and the transit depth at the probed pressure corresponds to 1\%. The region where atmospheres are hydrodynamically unstable has been left blank.  \textbf{Middle:} The same plot but 10\% abundance of the metallicity seen in WASP-39\,b. \textbf{Right:} The same plot but 1\% abundance of the metallicity seen in WASP-39\,b.  \label{fig:metallicity_radius_Temp} }
\end{figure*}

The significant difference in the pressure level probed in WASP-39~b's atmosphere and in the synthetic cases explored in N22 is primarily explained by the difference in the metallicity considered in two cases where a larger abundance of H$_{\rm 2}$O leads to the transmission spectrum being sensitive to lower pressure regimes. The pressure level predominantly probed at a given wavelength can be estimated following equations introduced in \citet{dewit2013}. The effective atmospheric height ($h_{\rm eff}(\lambda)$) corresponds to the altitude at which the optical depth $\tau(\lambda)$ is equal to $e^{-\gamma_{EM}}$, where $\gamma_{EM}$ is the Euler-Mascheroni constant \cite{Euler1740}. In the Voigt approximation and for an isothermal atmosphere (Eq.~S.29), 

 \begin{eqnarray}
\tau(z,\lambda) & = & \sqrt{2 \pi R_{p,0} H} \Lambda_{\kappa}\frac{2}{\sqrt{b_{\kappa}}} P_0 e^{-z/H},
 \label{tau_lines_T_cst_infl}
\end{eqnarray}
where $R_{p,0}$ is a planetary radius of reference, $H$ is the atmospheric scale height, $P_0$ is the reference pressure, and $\Lambda_{\kappa}$ and $b_{\kappa}$ are model parameters for the absorption coefficient--which depends on the molecule of interest, its abundance, as well as temperature and pressure. For most $\lambda$, $\Lambda_{\kappa}$ has the form of a product between the mixing ratio of the dominant opacity source and its attenuation coefficient at $\lambda$. Therefore, one can estimate the pressure level probed $P(h_{\rm eff},\lambda)$ via:
 \begin{eqnarray}
P(\rm h_{eff},\lambda) & = & \frac{2e^{-\gamma_{EM}}\sqrt{b_{\kappa}}}{\Lambda_{\kappa}\sqrt{2 \pi R_{p,0} H}}, \text{ and thus}\\
& \propto &  R_{p,0}^{-.5} H^{-.5} \Lambda_{\kappa}^{-1}.
 \label{eqn:tau_lines_T_cst_infl}
\end{eqnarray}
In other words, the bigger the planet ($R$), the puffier its atmosphere ($H$), and/or the stronger and more abundant its opacity source ($\Lambda_{\kappa}$), the shallower the atmospheric layers probed via transmission spectroscopy. As a result, large, hot, and high-metallicity planets such as WASP-39\,b are less affected by the opacity concerns introduced in N22. \autoref{fig:metallicity_radius_Temp} highlights that a significant SSD ($\geq5$) is found for small, low-metallicity, and/or cooler planets, many of which are already Cycle-1 targets for JWST (e.g., mini-Neptunes like K2-18\,b -- GO 2722, GJ 436 b -- GO 1185). The low metallicity relates to the opacity source, while the lower temperature decreases scale height ($H$) and pushes the transition from the Voigt regime to the Doppler regime to higher altitudes. 

It is worth noting that metallicity does not relate directly to opacity sources. Therefore high-metallicity atmospheres may be probed at high-pressure levels and can thus still be severely impacted by inadequate modeling of pressure-broadening effects. Earth's example shown in \autoref{fig:PT_Profile} highlights a case for which a high-metallicity driven by weak opacity sources (N$_2$, O$_2$, and Ar) does result in a contribution function peaking at high pressure levels.

\section{Conclusion}
\label{sec:conclusion}

We find that the main atmospheric properties (incl., abundance of main absorbers and p-T profile) of WASP-39~b can be retrieved from one JWST's NIRSpec/G395H observation with marginal effects stemming from current opacity models. The 0.5~dex opacity-driven accuracy wall reported in N22 does not apply in this case as the abundance of the dominant opacity sources such as H$_{\rm 2}$O and  CO$_{\rm 2}$ appears to be reliably constrained to within 0.30~dex. We explore the discrepancy between this work and N22 and find that the opacity-driven effects previously reported are marginal for atmospheres mostly probed below $<$1~mbar, which is typically the case for large, hot, and/or optically thick atmospheres. 

While planetary atmospheres with strong and abundant opacity sources (such as H$_{\rm 2}$O for WASP-39~b) will be reliably studied with current opacity models, many others will be limited by the accuracy wall introduced in N22. Cases such as the synthetic super-Earth and warm-Jupiter scenarios introduced in N22 or Earth--whose atmosphere is mostly probed in the $\sim$100 mbar regime (\autoref{fig:PT_Profile} \& \autoref{fig:metallicity_radius_Temp})--will remain challenging to study until data-driven improvements to the community's opacity models are implemented \citep{fortney2016}. Additionally, line list incompleteness is a universal issue that goes beyond WASP-39\,b retrievals. Lack of line lists that are complete at these hot atmospheric conditions can lead to significant biases (e.g. HITRAN vs HITEMP) to non-identification of the molecular features (e.g. see JTECERST23).

In the meantime, we advise future detailed atmospheric studies using transmission spectroscopy to be mindful of the reliability of their inferences when these reach below the $\sim$0.5 dex level. As highlighted in \citet{gharib2019}, emission spectroscopy and climate modelling probe/rely on deeper parts of a planetary atmosphere, and will therefore remains more susceptible to pressure related effects. In both cases, using SSD with different sets of cross-sections will allow for a quick assessment of the potential impact of opacities during the atmospheric retrievals. 

We note that significant differences are found between our retrieved abundances and results recently published  \citep{constantinou2023}. Such discrepancies may be due to differences in the data reduction, the inclusion of additional data from \textit{Hubble}, and/or the atmospheric models (incl., cloud and chemistry modeling). Such model-dependency investigation goes beyond the scope of this Letter, which aims to investigate the specific origin and extent of the opacity challenge for the atmospheric retrieval of gas giants with JWST. Future works by the Early Release Science team will provide the community with definitive and comprehensive insights into WASP-39~b's atmosphere, incl. important discussion regarding the aforementioned model sensitivities.


\vspace{5mm}
\facilities{JWST}

\software{\texttt{emcee} \citep{emcee}, \texttt{tierra} \citep{niraula2022_Nature} }

\acknowledgments
P. Niraula acknowledges \textbf{help from  general discussion on retrievals with Ana Glidden and group discussions led by Dr. Luis Welbanks}. This work is based on observations made with the NASA/ESA/CSA JWST. The data were obtained from the Mikulski Archive for Space Telescopes at the Space Telescope Science Institute, which is operated by the Association of Universities for Research in Astronomy, Inc., under NASA contract NAS 5-03127 for JWST. These observations are associated with program JWST-ERS-01366. Support for program JWST-ERS-01366 was provided by NASA through a grant from the Space Telescope Science Institute. The authors acknowledge the MIT SuperCloud and Lincoln Laboratory Supercomputing Center for providing (HPC, database, consultation) resources that have contributed to the research results reported within this paper.

\appendix

\begin{deluxetable*}{l|c|c|c|c|c}
\setlength{\tabcolsep}{5pt}
\tablecaption{\label{tab:retrieved_abundances_wasp39} Retrieved Abundances for different molecules with reference radius  1.25 R$_{\rm Jup}$ }
\tablehead{
\colhead{} & \colhead{} & \colhead{} & \colhead{} & \colhead{} & \colhead{} 
}
\startdata
\hline
\multicolumn{6}{c}{NIRSpec-G395H}\\
\hline
\multicolumn{1}{c}{Parameter} &  \multicolumn{1}{c}{CS-H2-25} & \multicolumn{1}{c}{CS-H2-500} & \multicolumn{1}{c}{CS-N2-25}& \multicolumn{1}{c}{CS-N2-500} & \multicolumn{1}{c}{CS-pRT} \\
\hline
\multicolumn{3}{c}{\textbf{Isothermal}}\\
$\chi^2$&1.08&1.08&1.07&1.07&1.08\\
SSD & - & 0.17 &  0.49 &  1.25 &  7.38 \\
PSD & - & 0.02 & 0.04 & 0.23 & 1.39 \\
AIC&380.36&380.25&378.10&377.65&383.12\\
BIC&422.61&422.5&420.35&419.9&425.37\\
Mass&0.265$^{+0.028}_{-0.029}$&0.266$^{+0.029}_{-0.030}$&0.267$^{+0.029}_{-0.029}$&0.267$^{+0.029}_{-0.029}$&0.271$^{+0.029}_{-0.030}$\\
Log P$_0$&-1.72$^{+0.18}_{-0.18}$&-1.73$^{+0.17}_{-0.18}$&-1.71$^{+0.18}_{-0.17}$&-1.75$^{+0.17}_{-0.16}$&-1.91$^{+0.17}_{-0.15}$\\
T$_0$&812.0$^{+47.0}_{-35.0}$&811.0$^{+49.0}_{-35.0}$&808.0$^{+40.0}_{-36.0}$&804.0$^{+38.0}_{-36.0}$&847.0$^{+42.0}_{-62.0}$\\
Log$_{\rm 10}$ [MR CH$_4$]&-5.55$^{+0.24}_{-0.27}$&-5.55$^{+0.24}_{-0.26}$&-5.57$^{+0.23}_{-0.28}$&-5.54$^{+0.22}_{-0.26}$&-5.55$^{+0.25}_{-0.31}$\\
Log$_{\rm 10}$ [MR CO]&-14.0$^{+8.0}_{-7.0}$&-14.0$^{+8.0}_{-7.0}$&-14.0$^{+8.0}_{-7.0}$&-14.0$^{+7.0}_{-7.0}$&-14.0$^{+8.0}_{-7.0}$\\
Log$_{\rm 10}$ [MR CO$_{\rm 2}$]&-2.99$^{+0.27}_{-0.27}$&-2.98$^{+0.26}_{-0.28}$&-2.99$^{+0.26}_{-0.28}$&-2.95$^{+0.24}_{-0.25}$&-3.13$^{+0.26}_{-0.26}$\\
Log$_{\rm 10}$ [MR H$_{\rm 2}$O]&-1.26$^{+0.18}_{-0.20}$&-1.26$^{+0.18}_{-0.21}$&-1.28$^{+0.18}_{-0.20}$&-1.23$^{+0.16}_{-0.19}$&-1.36$^{+0.19}_{-0.19}$\\
Log$_{\rm 10}$ [MR H$_{\rm 2}$S]&-1.70$^{+0.2}_{-0.24}$&-1.71$^{+0.2}_{-0.24}$&-1.71$^{+0.19}_{-0.22}$&-1.81$^{+0.21}_{-0.24}$&-1.65$^{+0.18}_{-0.23}$\\
Log$_{\rm 10}$ [MR HCN]&-15.0$^{+7.0}_{-7.0}$&-15.0$^{+7.0}_{-7.0}$&-15.0$^{+7.0}_{-7.0}$&-15.0$^{+7.0}_{-7.0}$&-15.0$^{+7.0}_{-7.0}$\\
Log$_{\rm 10}$ [MR SO$_{\rm 2}$]&-4.71$^{+0.27}_{-0.27}$&-4.72$^{+0.26}_{-0.26}$&-4.74$^{+0.26}_{-0.26}$&-4.70$^{+0.25}_{-0.26}$&-4.51$^{+0.24}_{-0.26}$\\
Log$_{\rm 10}$ [C/O]&-1.73$^{+0.15}_{-0.13}$&-1.73$^{+0.15}_{-0.13}$&-1.72$^{+0.14}_{-0.13}$&-1.72$^{+0.13}_{-0.12}$&-1.77$^{+0.13}_{-0.12}$\\
\hline
\multicolumn{3}{c}{\textbf{Parametric p-T Profile}}\\
$\chi^2$&1.03&1.03&1.02&1.02&1.03 \\
SSD & - & 0.48 &  0.56 &  2.04 &  7.70 \\
PSD & - & 0.72 & 0.22 & 0.86 & 0.75 \\
AIC & 366.54 & 367.21& 364.13& 364.7& 367.93 \\
BIC&416.47&417.14&414.06&414.63&417.86 \\
Mass&0.265$^{+0.027}_{-0.030}$&0.263$^{+0.030}_{-0.024}$&0.267$^{+0.028}_{-0.029}$&0.266$^{+0.028}_{-0.023}$&0.269$^{+0.027}_{-0.025}$\\
Log P$_0$&-2.15$^{+0.50}_{-0.25}$&-2.18$^{+0.58}_{-0.22}$&-2.16$^{+0.38}_{-0.21}$&-2.16$^{+0.31}_{-0.21}$&-2.29$^{+0.45}_{-0.21}$\\
T$_0$&2461.0$^{+412.0}_{-1669.0}$&2597.0$^{+296.0}_{-1821.0}$&2594.0$^{+301.0}_{-1626.0}$&2613.0$^{+286.0}_{-1697.0}$&2640.0$^{+270.0}_{-1845.0}$\\
$\sigma$&0.38$^{+0.19}_{-0.38}$&0.42$^{+0.18}_{-0.41}$&0.44$^{+0.19}_{-0.39}$&0.45$^{+0.21}_{-0.18}$&0.38$^{+0.13}_{-0.38}$\\
T$_\infty$&761.0$^{+71.0}_{-66.0}$&759.0$^{+56.0}_{-67.0}$&742.0$^{+56.0}_{-74.0}$&747.0$^{+56.0}_{-58.0}$&733.0$^{+110.0}_{-73.0}$\\
Log$_{\rm 10}$ [MR CH$_4$]&-5.81$^{+0.39}_{-9.1}$&-5.92$^{+0.57}_{-10.6}$&-5.98$^{+0.52}_{-9.3}$&-6.11$^{+0.63}_{-12.4}$&-10.0$^{+4.0}_{-10.0}$\\
Log$_{\rm 10}$ [MR CO]&-13.0$^{+7.0}_{-9.0}$&-15.0$^{+8.0}_{-7.0}$&-14.0$^{+7.0}_{-8.0}$&-15.0$^{+7.0}_{-7.0}$&-15.0$^{+7.0}_{-7.0}$\\
Log$_{\rm 10}$ [MR CO$_{\rm 2}$]&-3.05$^{+0.30}_{-0.28}$&-3.01$^{+0.30}_{-0.35}$&-3.03$^{+0.27}_{-0.34}$&-2.99$^{+0.32}_{-0.33}$&-3.26$^{+0.30}_{-0.34}$\\
Log$_{\rm 10}$ [MR H$_{\rm 2}$O]&-1.29$^{+0.21}_{-0.25}$&-1.28$^{+0.27}_{-0.30}$&-1.31$^{+0.24}_{-0.30}$&-1.29$^{+0.23}_{-0.28}$&-1.48$^{+0.25}_{-0.32}$\\
Log$_{\rm 10}$ [MR H$_{\rm 2}$S]&-1.90$^{+0.28}_{-0.36}$&-1.97$^{+0.29}_{-0.37}$&-2.00$^{+0.31}_{-0.36}$&-2.06$^{+0.31}_{-0.48}$&-2.1$^{+0.44}_{-0.49}$\\
Log$_{\rm 10}$ [MR HCN]&-16.0$^{+6.0}_{-7.0}$&-16.0$^{+6.0}_{-6.0}$&-16.0$^{+7.0}_{-6.0}$&-14.0$^{+5.0}_{-8.0}$&-15.0$^{+6.0}_{-7.0}$\\
Log$_{\rm 10}$ [MR SO$_{\rm 2}$]&-4.51$^{+0.27}_{-0.32}$&-4.49$^{+0.26}_{-0.26}$&-4.54$^{+0.26}_{-0.27}$&-4.52$^{+0.25}_{-0.25}$&-4.47$^{+0.27}_{-0.24}$\\
Log$_{\rm 10}$ [C/O]&-1.76$^{+0.13}_{-0.12}$&-1.74$^{+0.12}_{-0.13}$&-1.74$^{+0.13}_{-0.12}$&-1.71$^{+0.11}_{-0.14}$&-1.79$^{+0.12}_{-0.12}$\\
\enddata
\tablenotetext{}{}
\end{deluxetable*}

\bibliography{Bibliography}{}
\end{document}